\documentclass[lettersize,journal]{IEEEtran}
\usepackage{amsmath,amssymb,amsfonts}
\usepackage{graphicx}
\usepackage{float}
\usepackage[caption=false,font=normalsize,labelfont=sf,textfont=sf]{subfig}
\usepackage{color}
\usepackage{cite}
\usepackage{multirow}
\usepackage{algorithm}  
\usepackage{algorithmicx}  
\usepackage{algpseudocode} 
\usepackage{bm}
\usepackage[colorlinks,linkcolor=black,anchorcolor=black,
citecolor=black,urlcolor=black]{hyperref}

\usepackage{stfloats}

\usepackage{siunitx}
\DeclareSIUnit\bps{bps}
\DeclareSIUnit\Torr{Torr}
\DeclareSIUnit\torr{Torr}
\DeclareSIUnit\sample{Sa}
\newcommand*{\circled}[1]{\lower.7ex\hbox{\tikz\draw (0pt, 0pt)%
  circle (.5em) node {\makebox[1em][c]{\small #1}};}}

\ifCLASSINFOpdf

\else

\fi

\graphicspath{{figures/}}

\begin{document}

\title{Channel Measurement and Coverage Analysis for NIRS-Aided THz Communications in Indoor Environments}

\author{Yuanbo~Li, Yiqin Wang, Yi Chen, Ziming Yu, and Chong~Han,~\IEEEmembership{Member,~IEEE}
\thanks{
Yuanbo Li, Yiqin Wang, and Chong Han are with the Terahertz Wireless Communications (TWC) Laboratory, Shanghai Jiao Tong University, Shanghai, China (e-mail: \{yuanbo.li, wangyiqin, chong.han\}@sjtu.edu.cn).

Yi Chen and Ziming Yu are with Huawei Technologies Co., Ltd, Chengdu, China (e-mail: \{chenyi171, yuziming\}@huawei.com).
}
}

	{}
	\maketitle
	\thispagestyle{empty}

\begin{abstract}
Due to large reflection and diffraction losses in the THz band, it is arguable to achieve reliable links in the none-line-of-sight (NLoS) cases. Intelligent reflecting surfaces, although are expected to solve the blockage problem and enhance the system connectivity, suffer from fabrication difficulty and operation complexity. 
In this work, non-intelligent reflecting surfaces (NIRS), which are simply made of costless metal foils and have no signal configuration capability, are adopted to enhance the signal strength and coverage in the THz band. Channel measurements are conducted in typical indoor scenarios at 306-321 GHz and 356-371 GHz bands to validate the effectiveness of the NIRS. Results measured with NIRS in different sizes show that large NIRS performs much better than small NIRS. Furthermore, by invoking the NIRS, the additional reflection loss can be reduced by more than 10~dB and the coverage ratio is increased by up to 39$\%$ for a 10~dB signal-to-noise ratio (SNR) threshold.
\end{abstract}
\section{Introduction}
\IEEEPARstart{T}{he} THz band, ranging from \SI{0.1}{THz} to \SI{10}{THz}, is envisoned as key technology for the next-generation communication networks, stemming from abundant spectrum resources and ultra-large contiguous usable bandwidths~\cite{rappaport2019wireless}. However, the coverage ability of THz communications is a concern, especially in none-line-of-sight (NLoS) cases, due to larger reflection and diffraction losses, as well as weaker penetration abilities~\cite{wu2021interference,han2015multiray,akyildiz2018combating,Jacob2012Diffraction,eckhardt2021channel}. As a result, when line-of-sight (LoS) transmission is not available, the received power could be reduced drastically where reliable THz communications might be hard to achieve. 
\par To improve the coverage situation in NLoS region, intelligent reflecting surface (IRS) is recently studied. IRS is a kind of tunable metasurface consisting of a large number of reflecting elements, which are able to manipulate the reflecting amplitude and phase shift of the impinging THz waves~\cite{9690477,9326394,Bjornson2020Reconfigurable}. As a result, the THz waves can be redirected and the coverage ability of THz communications is possibly much improved. {However, the flexibility of IRS comes at a price. First, thousands of elements are required to compensate for the path loss of IRS-aided THz communications, since the IRS elements scatter the radio signal as if from a point source~\cite{8910627,Ozdogan2020Intelligent}. It is difficult and costly to fabricate and control such large amount of IRS elements. Second, the effective usage of IRS needs to address many design challenges, such as passive beamforming algorithms, IRS channel acquisition, etc. }
\par {In contrast, an alternative scheme is non-intelligent reflecting surface (NIRS), which is a non-tunable surface made by materials that have small reflection and scattering losses, such as aluminium foils~\cite{abbasi2021ultra}. Compared to IRS, NIRS is has almost no cost, no fabrication, and super-easy deployment.} Even though several studies have revealed the effectiveness of NIRS in mmWave bands~\cite{Khawaja2020Coverage,El2022Enhancement,Maeng2022Coverage}, according to authors' best knowledge, the efficacy of NIRS for THz communications have not been fully analyzed. We attempt to fill this research gap, based on measurements.
\par In this letter, measurement campaigns are conducted in indoor corridor and hallway scenarios at \SIrange{306}{321}{GHz} and \SIrange{356}{371}{GHz} in the NLoS case. In all scenarios and bands, the THz channels are measured twice with the same layout, where the NIRS is included or excluded. For comparison, measurements are conducted with small NIRS deployed in different locations. Results reveal that the performance of small NIRS is worse than those of large NIRS and highly dependent on deploying locations. Therefore, it is better to use large NIRS for good performance. Furthermore, the additional reflection loss and coverage situations are thoroughly analyzed and compared. Results have shown that by using the NIRS,the additional reflection loss can be reduced by more than 10~dB and the coverage ratio in the NLoS area is increased by up to 39$\%$ for a 10~dB signal-to-noise ratio (SNR) threshold. 
\par The remainder of this letter is organized as follows. In Sec.~\ref{sec:measurement}, the measurement campaigns are introduced in detail. Moreover, the data processing procedure is introduced in Sec.~\ref{sec:processing}. Furthermore, the measurement results are carefully elaborated in Sec.~\ref{sec:results}. Finally, Sec.~\ref{sec:conclude} concludes the letter.
\section{Channel Measurement Campaign}
\label{sec:measurement}
\par In this section, the measurement campaigns for NIRS-aided communications are described, including the sounder system, the measurement setup and the measurement deployment.
\subsection{Sounder System}
\par To measure the THz channels, a vector network analyzer (VNA)-based channel sounder is used, whose measurable bands ranges from \SI{260}{GHz} to \SI{400}{GHz}. Furthermore, both transmitter (Tx) and receiver (Rx) modules are installed on electric carts, lifters and rotators, which support us to conveniently change the locations, heights and pointing directions. For full-fledged description of our measurement system, readers are encouraged to refer to~\cite{li2022channel,wang2022thz}.
\subsection{Measurement Setup}









        
        





{Two frequency bands are measured, namely \SIrange{306}{321}{GHz} and \SIrange{356}{371}{GHz}, across a \SI{15}{GHz} wide band.} The channel transfer functions (CTFs) are measured with a \SI{2.5}{MHz} sweeping interval. Therefore, the maximum delay of multi-path components are \SI{400}{ns}, corresponding to a maximum path length of \SI{120}{m}. The time resolution is \SI{66.7}{ps}. The heights of Tx and Rx are \SI{2}{m} and \SI{1.75}{m}, respectively. Moreover, the transmitter is equipped with a standard waveguide WR2.8, which has \SI{7}{dBi} antenna gain and a $30^\circ$ half-power beamwidth (HPBW). By contrast, the Rx is equipped with a directional antenna with a \SI{25}{dBi} antenna gain and a $8^\circ$ HPBW. To evaluate received power from various directions, the Rx scans the spatial domain with $10^\circ$ angle steps, from $0^\circ$ to $360^\circ$ in the azimuth plane and $-20^\circ$ to $20^\circ$ in the elevation plane.
\subsection{Measurement Deployment}
\begin{figure}[!tbp]
    \centering
    \subfloat[] {
     \label{fig:pic_cor}     
    \includegraphics[width=0.76\columnwidth]{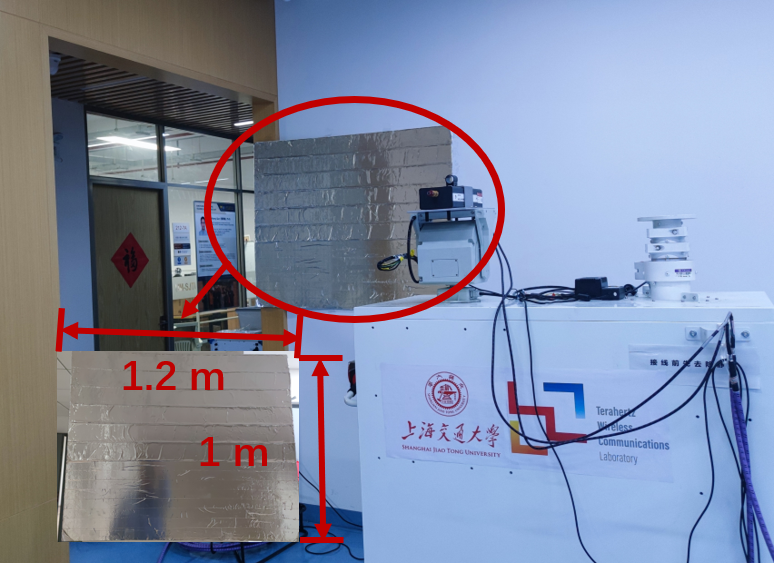}  
    }
    \quad
    \subfloat[] {
     \label{fig:deploy_cor}     
    \includegraphics[width=0.8\columnwidth]{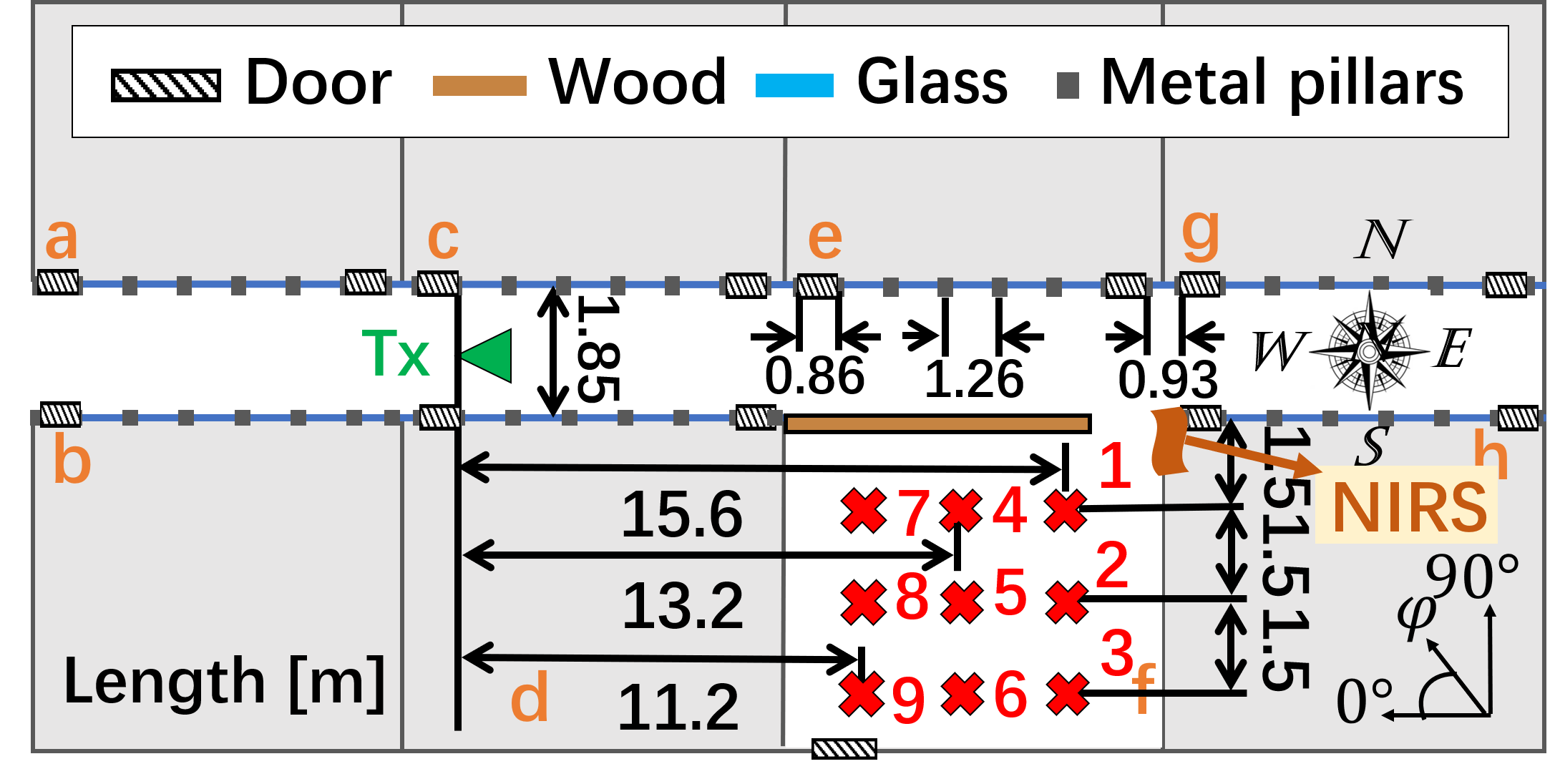}  
    }   
    \caption{The measurement in the corridor. (a) Pictures. (b) Bird's eye view.}
    \label{fig:corridor}
    \vspace{-0.5cm}
\end{figure}
\begin{figure}[!tbp]
    \centering
    \subfloat[] {
     \label{fig:pic_hallway}     
    \includegraphics[width=0.76\columnwidth]{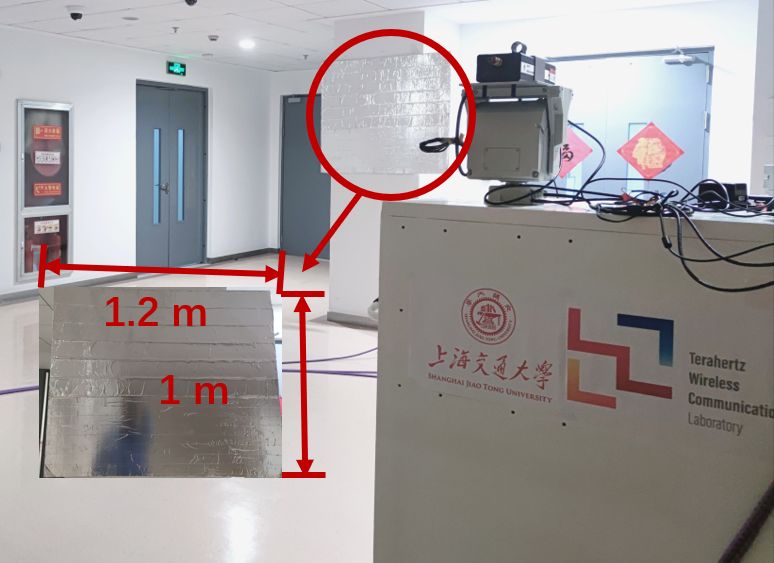}  
    }
    \quad
    \subfloat[] {
     \label{fig:deploy_hallway}     
    \includegraphics[width=0.72\columnwidth]{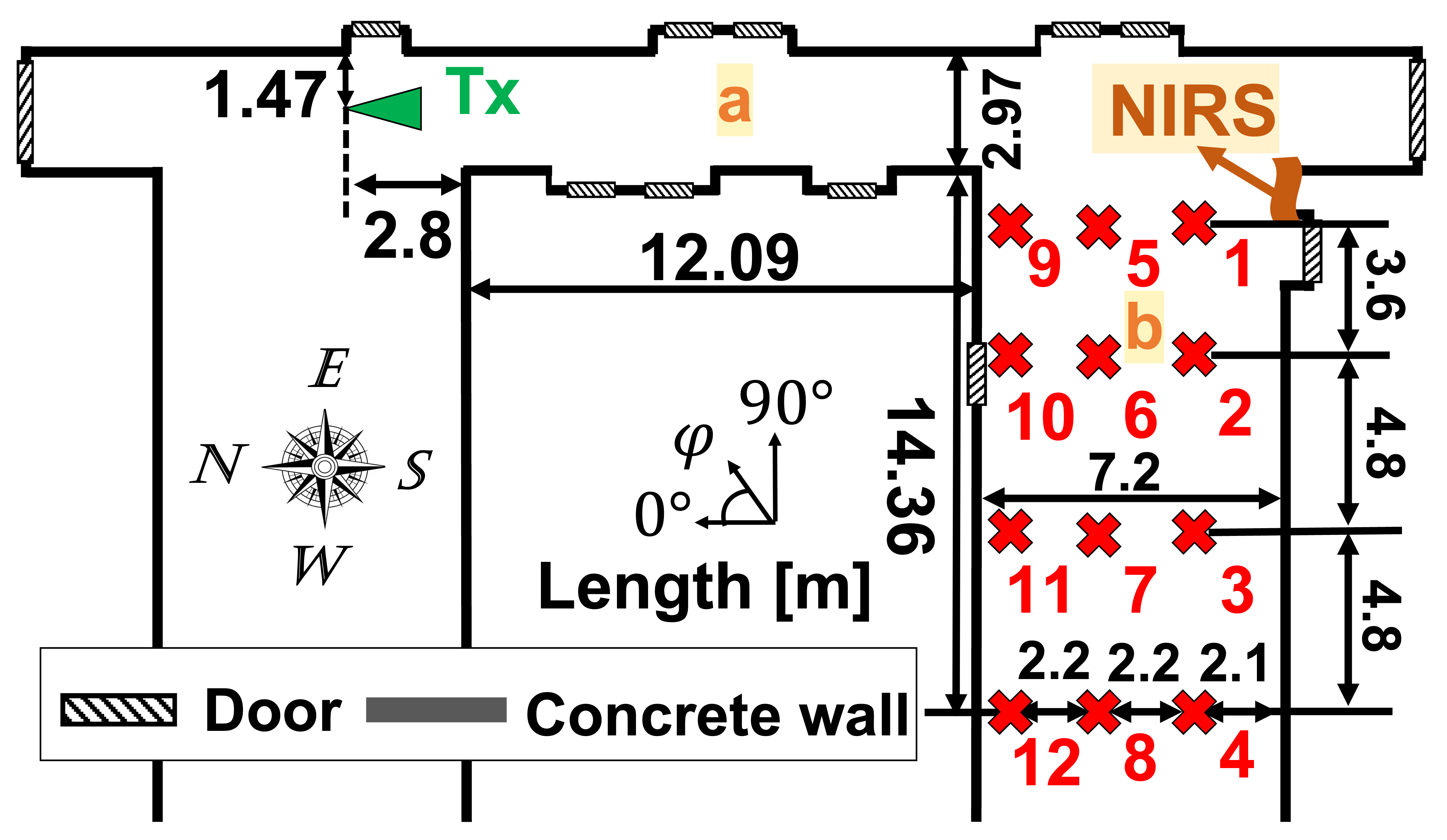}  
    }   
    \caption{The measurement in the hallway. (a) Pictures. (b) Bird's eye view.}
    \label{fig:hallway}
    \vspace{-0.5cm}
\end{figure}
\par The measured scenarios are shown in Fig.~\ref{fig:corridor} and Fig.~\ref{fig:hallway}. In the corridor scenario, the transmitter is placed near the room c in the corridor and remains static, while 9 Rx positions are deployed in room f, whose positions are marked in Fig.~\ref{fig:corridor} (b). Moreover, the NIRS are glued around the turning corner, as shown in Fig.~\ref{fig:corridor} (a). The NIRS is made of aluminium foils and has a size around \SI{1.2}{m}$\times$\SI{1}{m}.  Similarly, in the hallway scenario, the transmitter locates around the north end of the hallway a, while 12 Rx positions are selected in hallway b in the NLoS case. The NIRS with a size of \SI{1.2}{m}$\times$\SI{1}{m} is glued near the turning corner, on the southern wall of hallway b. 
Note that the NIRS is glued on walls rather than being placed in the hallway/corridor area to avoid blocking any area that is already in LoS.
\par {To investigate how the size and locations of NIRS affect reflecting and thus converge performance, measurements are conducted with a small NIRS covering only one ninth area of the large NIRS mentioned above. In the first measurement, the small NIRS is deployed at the top-left corner of the area where large NIRS is deployed, followed by varying the small NIRS from left to right, and from top to bottom, for nine different locations. Due to high time consumption of channel measurements, only Rx point 1 in the hallway scenario is measured with small NIRS.}
\section{Data Processing Procedure}
\label{sec:processing}
\par In this section, the data processing procedure is explained, including the calibration of raw data and noise elimination in power-delay-angular profile (PDAP).
\subsection{Calibration of Raw Data}
\par To eliminate unwanted factors, such as effects of cables, amplifiers, etc., the measured raw data is calibrated. Firstly, S parameters are measured when Tx and Rx are directly-connected through waveguides, which only includes the unwanted factors. Then, the Tx/Rx are located in certain positions and real measurements are conducted, where the measured results include both the system effects and the CTFs of THz channels. As a result, the CTF of the THz channel can be expressed as,
\begin{equation}
    H=\frac{S_{21}^{\text{measure}}}{S_{21}^{\text{extra}}S_{21}^{\text{connect}}}
\end{equation}
where $S_{21}^{\text{extra}}$ represents influences of components due to the different set-ups when conducting real measurement and directly-connected measurement. For example, horn antennas are only used in the real measurement.
\subsection{Noise Elimination in PDAP}
\par Based on the CTFs, the channel impulse responses (CIRs) of the THz channel can be obtained through inverse Fourier transform, i.e., $h=\mathcal{F}^{-1}(H)$. Furthermore, by merging CIRs in different scanning directions into a whole matrix, the PDAP can be obtained, as
\begin{equation}
    P_{i,j,k} [\text{dB}]=20\log_{10}|h_{i,j}[k]|
\end{equation}
where $h_{i,j}[k]$ denote the CIR at $k^\text{th}$ temporal sample in $i^\text{th}$ elevation scanning direction and $j^\text{th}$ azimuth scanning direction.
\par To avoid the influences of noises, a threshold-based noise elimination method is used. Specifically, any PDAP samples with power lower than a certain threshold are treated as noise, which are then assigned with a \SI{-300}{dB} value, i.e., nearly zero value. The threshold is set as \SI{-160}{dB}, which is slightly higher than the maximum power of the PDAP samples measured when there is no signal transmitting.
\section{Measurement Results and Analysis}
\label{sec:results}
\par In this section, the measurement results are analyzed and discussed. To begin with, the influences of size and location of NIRS are studied. Moreover, based on directional path loss results, the additional reflection loss on NIRS/walls are calculated and compared. Last but not least, the coverage enhancement using the NIRS is analyzed.
\subsection{Influences of Size and Location of NIRS}
\begin{figure}
    \centering
    \includegraphics[width=0.9\columnwidth]{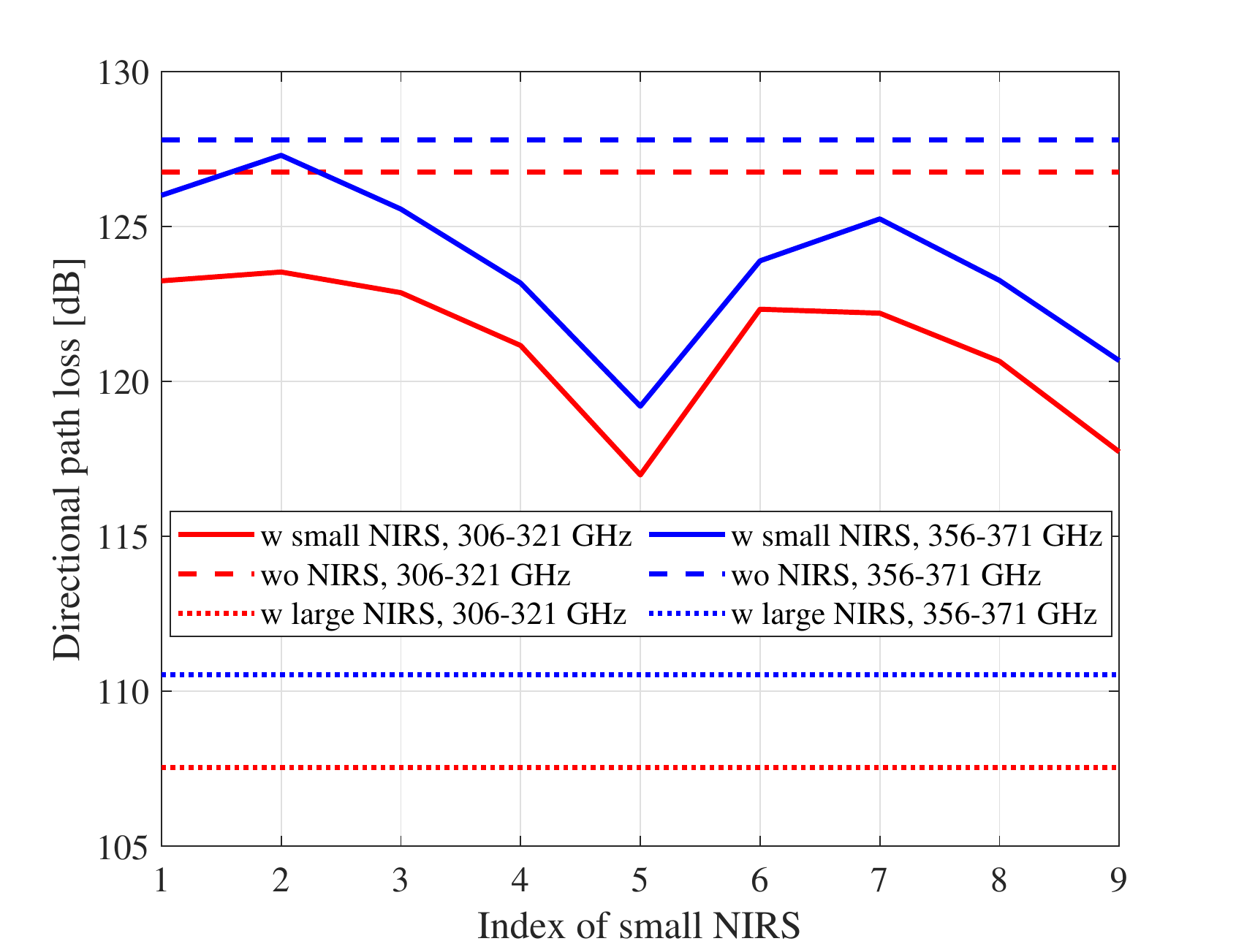}
    \caption{{Directional path loss measured with different NIRS deployment.}}
    \label{fig:smallNIRS}
    \vspace{-0.5cm}
\end{figure} 
\par {Since the NIRS is made of metal materials with smaller reflection loss compared to concrete walls, signals reflected on NIRS are expected to be enhanced. To prove this, the directional path loss is studied, calculated by summing signal power received from the NIRS-to-Rx direction, as
\begin{equation}
    \text{PL}_{\text{Dir}}~[\text{dB}]=-10\log_{10}(\sum_{i,j\in \bm{S}}\sum_{k}10^{P_{i,j,k}/10})
\end{equation}
where $\bm{S}$ contains the angle indexes of direction-of-arrival (DoA) that intersects with the NIRS, such as $140^\circ\sim170^\circ$ for Rx point 1 in the corridor.}
\par  {The directional path loss results with different sizes of NIRS are shown in Fig.~\ref{fig:smallNIRS}, from which several observations are made as follows. First, the directional path loss with small NIRS is within the limits of path losses with/without large NIRS, which indicates the small NIRS can enhance the received power, yet weaker than the large NIRS. Second, the directional path loss is highly dependent on the location of small NIRSs. Specifically, smaller path loss occurs at location 5 and 9, i.e., the middle and bottom-right areas of the NIRS, while other locations incur with larger path loss. Therefore, being insensitive to exact locations, large NIRS is a more effective choice to enhance the received power.}
\subsection{Additional Reflection Loss With/Without NIRS}
\par {The Tx-NIRS/wall-Rx link can be regarded as a virtual LoS link with additional reflection loss, whose path loss can be modeled by
\begin{equation}
    \text{PL}_\text{Dir}(f,d_1, d_2)=\text{PL}^\text{CI}_{\text{LoS}}(f,d_1+d_2)+\text{L}_\text{Ref}~[\text{dB}]
\end{equation}
where $\text{PL}^\text{CI}_\text{LoS}(f,d)$ is the close-in free space reference distance (CI) model, as reported in our previous work with PLEs of 1.39 in hallway scenario and 1.35 in corridor scenario~\cite{li2022channel,wang2022thz}. Additionally, $f$ is the carrier frequency. $d_1$ is the distance between Tx and NIRS and $d_2$ is the distance between NIRS and Rx. Moreover, $\text{L}_\text{Ref}$ denotes the additional reflection loss on NIRS/walls. Note that this additional reflection loss is a link-level characteristic, different from the frequently mentioned reflection loss, which evaluates how electromagnetic wave attenuates after being reflected on certain materials.}
\begin{figure}[!tbp]
    \centering
    \subfloat[]{ 
    \includegraphics[width=0.81\columnwidth]{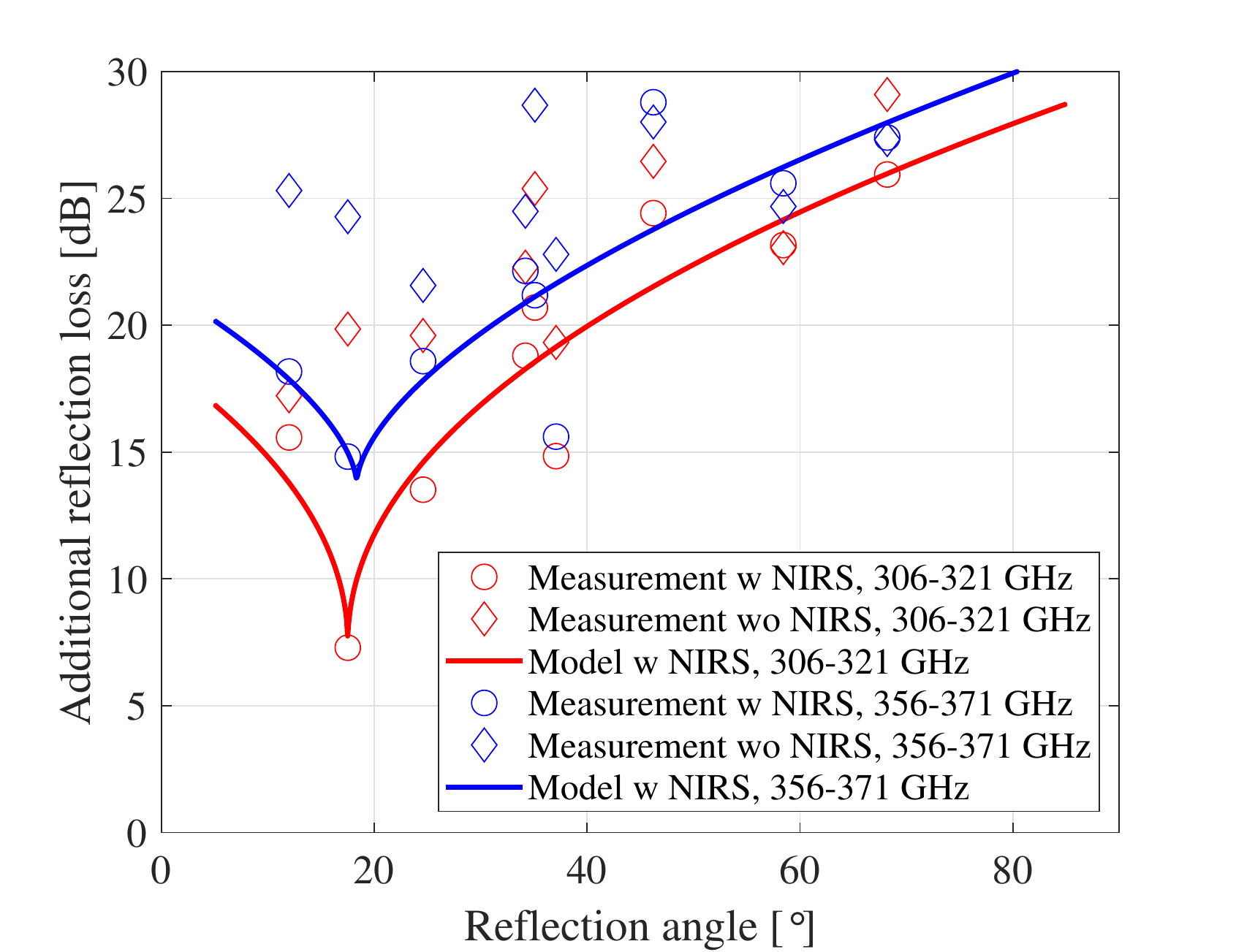}  
    }
    \vspace{-0.3cm}
    \quad
    \subfloat[]{ 
    \includegraphics[width=0.81\columnwidth]{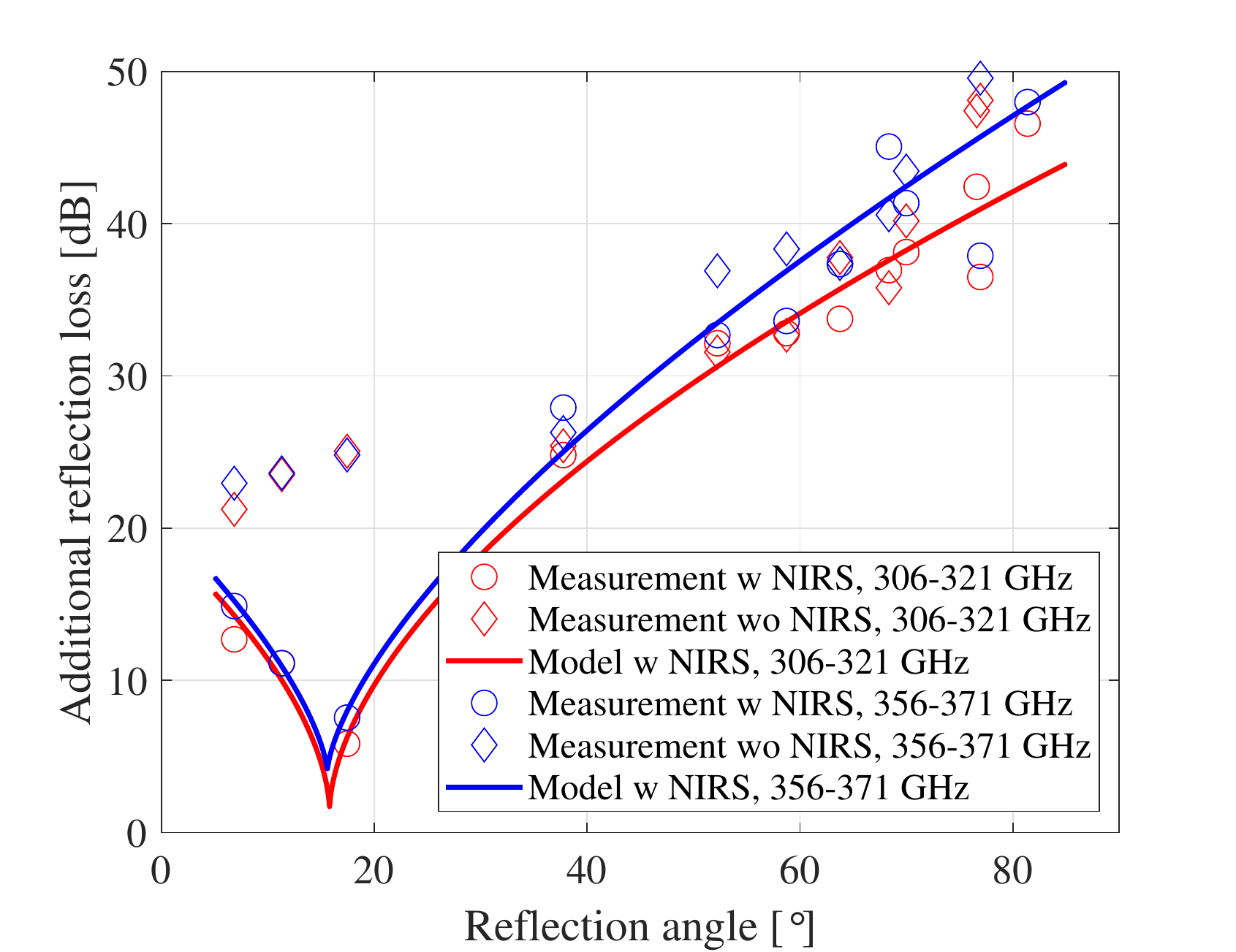}  
    }
    \caption{{The additional reflection loss results, in a (a) corridor and (b) hallway.}}
    \label{fig:refLoss}
    \vspace{-0.5cm}
\end{figure}
\par {The measured results of additional reflection loss are shown in Fig.~\ref{fig:refLoss}, where the results with NIRS are further fitted by a polynomial function with respect to the reflection angle, as
\begin{equation}
    \text{L}_\text{Ref}^\text{Pol} (\varphi)=a|\varphi-\overline{\varphi}|^b+c
\end{equation}
where $\varphi$ stands for the reflection angle, which is defined as the angle between the reflected direction and the normal direction of NIRS in the azimuth plane. $\overline{\varphi}$ is the reflection angle with minimum reflection loss. Moreover, $a$, $b$, and $c$ are fitting parameters, whose values are summarized in Table~\ref{tab:parameters}.}
\begin{table}[tbp]
	\centering 
	\caption{{Fitting parameters of additional reflection loss model.}} 
	\label{tab:parameters}  
	\begin{tabular}{|c|c|c|c|c|}  
		\hline  
		&\multicolumn{4}{c|}{}\\[-6pt]  
		\multirow{3}{*}{Fitting parameters} & \multicolumn{4}{c|}{Frequency bands}\\
		\cline{2-5}
            &\multicolumn{2}{c|}{}&\multicolumn{2}{c|}{}\\[-6pt]
                                                & \multicolumn{2}{c|}{306-321 GHz} & \multicolumn{2}{c|}{356-371 GHz}\\
		\cline{2-5}
            &&&&\\[-6pt]
                                                &Corridor&Hallway&Corridor&Hallway\\
            \hline
            &&&&\\[-6pt]
            $\overline{\varphi} [\circ]$&17.51&15.79&18.34&15.58\\
            \hline
            &&&&\\[-6pt]
            $a$&2.80&3.52&1.34&2.60\\
            \hline
            &&&&\\[-6pt]
            $b$&0.48&0.59&0.60&0.67\\
            \hline
            &&&&\\[-6pt]
            $c$&7.4&1.51&13.78&4.01\\
            \hline
                                                
	\end{tabular}
	\vspace{-0.5cm}
\end{table}
\par {First, polynomial model fits the measurement data very well, which proves its effectiveness. Second, the additional reflection loss with NIRS is highly dependent on the reflection angle, while those without NIRS is less regular. As the reflection angle deviates from the directions with minimum loss, namely $18^\circ$ in corridor and $16^\circ$ in hallway, the additional reflection loss increases. Third, for those reflection angles near the best directions, the additional reflection loss are greatly reduced compared to results without NIRS, by \SIrange{7}{19}{dB}, while the loss decrease in other directions are around \SI{3}{dB}. Compared to existing studies, the power enhancement measured in this work is smaller than those measured in~\cite{Khawaja2020Coverage} yet larger than those in~\cite{abbasi2021ultra}, due to the following reasons. On one hand, polished flat reflectors are used and their orientation are adjusted to provide strong specular reflections in~\cite{Khawaja2020Coverage}, while the reflectors in our work is rough and glued on walls, resulting in non-specular reflections with larger reflection loss. On the other hand, our measurements are conducted in corridor or hallway scenarios, where not only once-reflection but also higher order reflections could have significant power. In contrast, measurements in~\cite{abbasi2021ultra} are conducted in outdoor scenarios with fewer scatters, resulting in lower multipath richness and worse performance of NIRS.}
\par {Furthermore, the additional reflection loss shows different patterns in different scenarios and frequency bands. First, larger losses are observed at \SIrange{356}{371}{GHz}, compared to those at \SIrange{306}{321}{GHz}.} Moreover, different results are observed for different scenarios. On one hand, in the hallway scenario, the minimum additional reflection loss and corresponding reflection angles appear smaller than those in the corridor, since the hallway is wider than the corridor and a larger area of the NIRS is visible to the transmitter. On the other hand, larger losses are observed for Rx locations with large reflection angles ($>60^\circ$) in the hallway scenario, due to longer distance from the turning corner.
\subsection{Coverage Ratio With/Without NIRS}
\begin{figure}[!tbp]
    \centering
    \subfloat[]{ 
    \includegraphics[width=0.8\columnwidth]{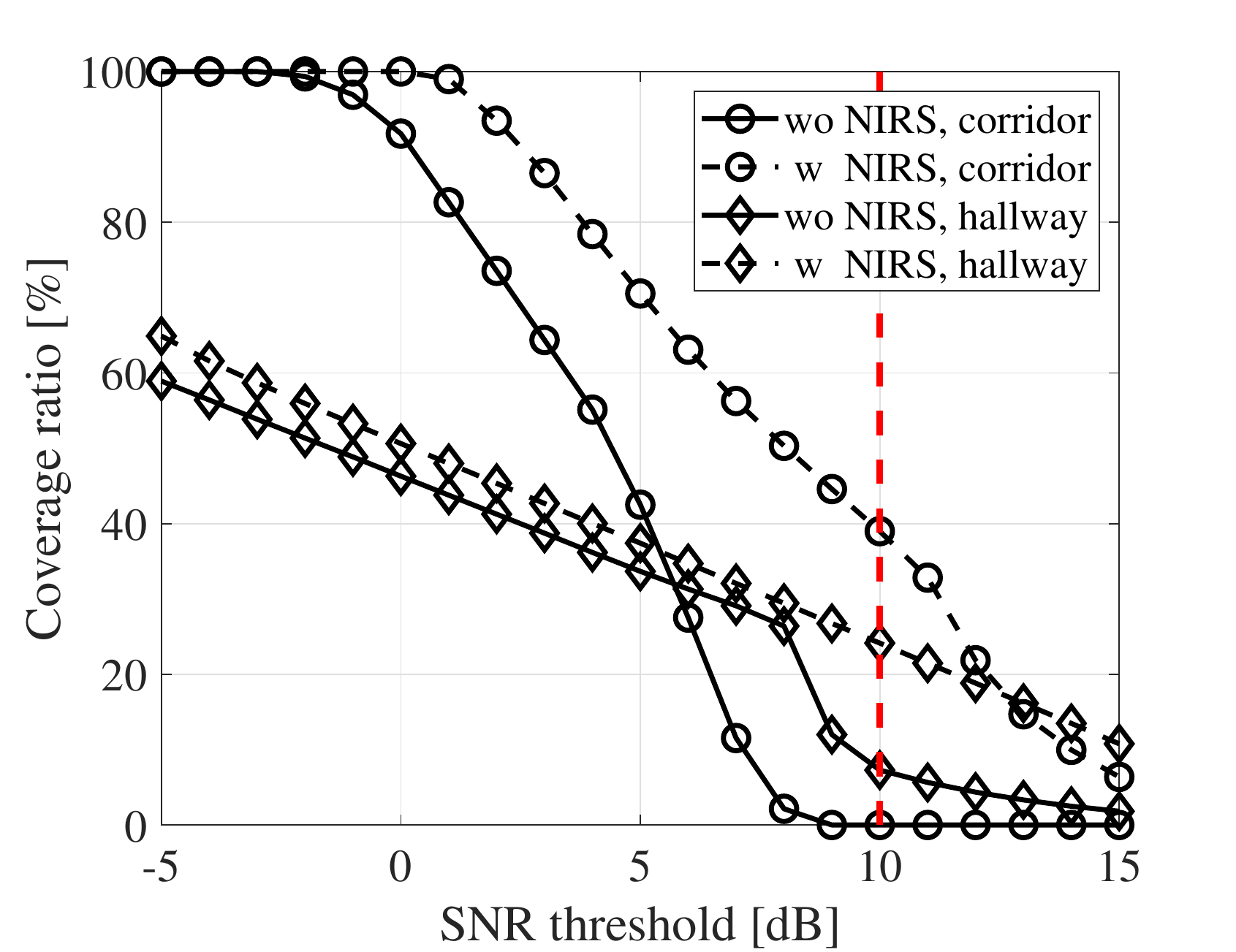}  
    }
    \vspace{-0.3cm}
    \quad
    \subfloat[]{ 
    \includegraphics[width=0.8\columnwidth]{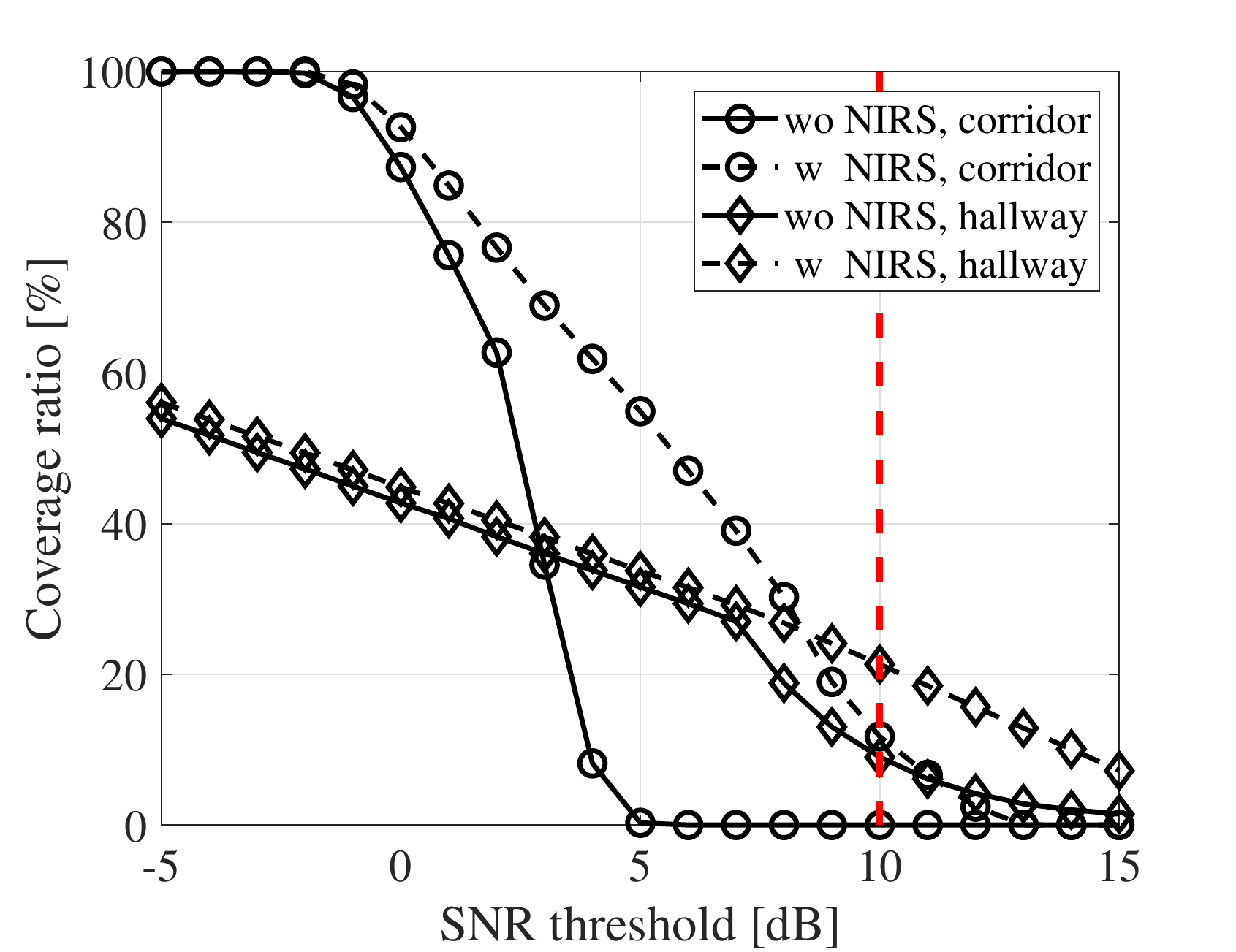}  
    }
    \caption{{The coverage ratio results, at (a) 306-321 GHz and (b) 356-371 GHz.}}
    \label{fig:coverage}
    \vspace{-0.5cm}
\end{figure}
\par To study the coverage situation in NLoS region, the omnidirectional path loss is calculated by summing power from all directions together, as
\begin{equation}
    \text{PL}_{\text{Omni}}~[\text{dB}]=-10\log_{10}(\sum_{i,j,k}10^{P_{i,j,k}/10})
\end{equation}
\par Due to high time consumption of channel measurements, only limited Rx positions are measured. Therefore, to analyze the coverage situations in the whole area, the omnidirectional path loss in positions between adjacent Rx locations are obtained through linear interpolation. The coverage ratio is defined as the percentage of the area where the received SNR is larger than a certain threshold. Specifically, the received SNR is calculated as
\begin{equation}
    \gamma~[\text{dB}]=\frac{P_tG_tG_r10^{-\text{PL}_\text{Omni}/10}}{FkTB}
\end{equation}
where $P_t$ is the transmit power. $G_{t/r}$ are the antenna gains of Tx/Rx. Moreover, $F$ is the noise figure of the receiver. Additionally, $k$ is the Boltzmann constant as 1.381$\times10^{-23}~\text{W}\cdot\text{S}/\text{K}$. $T$ and $B$ are the temperature and bandwidth, respectively.
\par According to the reference values in~\cite{Rikkinen2020thz}, the parameters of realistic THz communication links are considered as follows. The transmit power is \SI{13}{dBm}. Moreover, the temperature and bandwidth are set as \SI{300}{K} and \SI{15}{GHz}, which are consistent with our measurements. Additionally, the antenna gains of Tx and Rx are selected as \SI{25}{dBi}. The noise figure is set as \SI{10}{dB}.
\par The coverage ratios in indoor environments are shown in Fig.~\ref{fig:coverage}, from which we can make several observations as follows. First, without NIRS, the coverage in the corridor is better than that in the hallway with low SNR thresholds ($<$\SI{5}{dB}), while opposite situation occurs with high SNR thresholds. The reason is that in the corridor, significant scattering from metal pillars results in wide coverage with low SNR. However, since the turning corner in the hallway scenario is more open, there are more area in the hallway that receives high SNR ($>$\SI{5}{dB}). {Second, comparing different frequency bands, the coverage ratio at \SIrange{306}{321}{GHz} is better than that at \SIrange{356}{371}{GHz}, since higher frequencies result in a larger path loss. Third, by including the NIRS, the coverage ratio is significantly improved, especially with a high SNR threshold. Specifically, given a \SI{10}{dB} SNR threshold, the coverage ratio in the hallway is increased by $17\%$ and $12\%$ at \SIrange{306}{321}{GHz} and \SIrange{356}{371}{GHz}, respectively, while in the corridor, the coverage ratio is increased from zero percent (no coverage at all) to $39\%$ and $12\%$ at \SIrange{306}{321}{GHz} and \SIrange{356}{371}{GHz}, respectively.}
\subsection{Comparison With IRS}
\begin{table}[]
    \centering
    \caption{{Comparison of IRS and NIRS.}}
        \includegraphics[width=0.9\columnwidth]{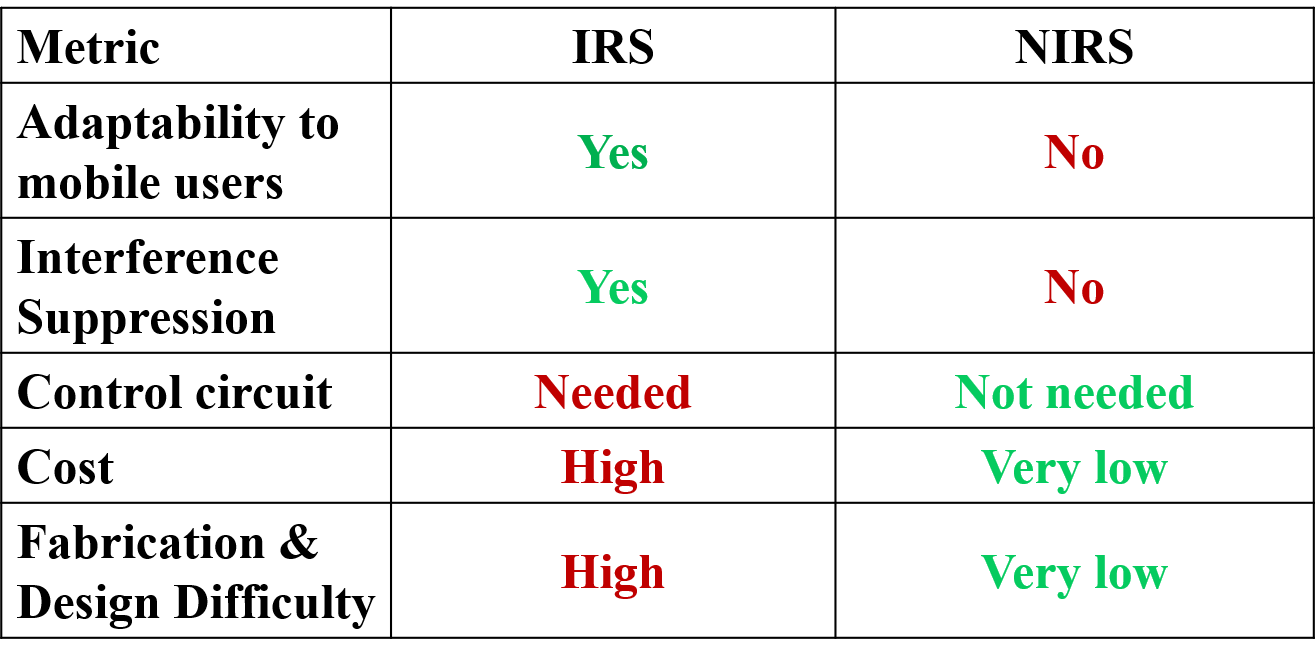}
    \label{tab:comparison}
    \vspace{-0.5cm}
\end{table}
\par {As listed in Table~\ref{tab:comparison}, both the IRS and NIRS have advantages and disadvantages. On one hand, IRS is much more flexible than NIRS. By adjusting the amplitude and phase of reflected signals on each element, IRS can achieve passive beamforming to adapt to mobile users and is also able to suppress interference from neighboring base stations. In contrast, NIRS has no such abilities. On the other hand, the fabrication and usage of NIRS is substantially cheaper and simpler, since it is made by costless metal foils and does not need control circuits. Since thousands of elements are needed in the THz band to compensate for large path losses, the fabrication and usage of IRS are much more costly and complicated in practice.}
\section{Conclusion}
\label{sec:conclude}
In this letter, measurement campaigns in typical indoor scenarios are conducted to testify the effectiveness of non-intelligent reflecting surface (NIRS). 9 Rx positions in the corridor scenario and 12 Rx positions in the hallway scenario are measured. Based on the measured data, the influences of size and location of NIRS, additional reflection loss, as well as coverage ratio are studied. Results have shown that it is better to use a large NIRS to enhance the received power. Moreover, the additional reflection loss can be reduced by more than \SI{10}{dB} with NIRS. Finally, even though the power enhancement is not uniform and hard to control, the overall coverage ratio could be improved by up to 39$\%$ by adding the NIRS. In general, NIRS offers a convenient way to improve the coverage of THz communications.
\bibliographystyle{IEEEtran}
\bibliography{IEEEabrv,main}
\end{document}